\documentclass[twocolumn,pre,superscriptaddress]{revtex4-2}

\usepackage{amssymb}
\usepackage{amsmath}
\usepackage{graphicx}
\usepackage{dcolumn}
\usepackage{bm}
\usepackage{epstopdf}
\usepackage{color}

\begin{document}

\title{Universal cover-time distribution of heterogeneous random walks}

\author{Jia-Qi Dong}
\affiliation{Lanzhou Center for Theoretical Physics and Key Laboratory of Theoretical Physics of Gansu Province, Lanzhou University, Lanzhou, Gansu 730000, China}
\affiliation{CAS Key Laboratory of Theoretical Physics, Institute of Theoretical Physics, CAS, Beijing 100190, China}

\author{Wen-Hui Han}
\affiliation{Lanzhou Center for Theoretical Physics and Key Laboratory of Theoretical Physics of Gansu Province, Lanzhou University, Lanzhou, Gansu 730000, China}

\author{Yisen Wang}
\affiliation{Lanzhou Center for Theoretical Physics and Key Laboratory of Theoretical Physics of Gansu Province, Lanzhou University, Lanzhou, Gansu 730000, China}

\author{Xiao-Song Chen}
\affiliation{School of Systems Science, Beijing Normal University, Beijing 100875, China}

\author{Liang Huang}
\thanks{Corresponding author: huangl@lzu.edu.cn}
\affiliation{Lanzhou Center for Theoretical Physics and Key Laboratory of Theoretical Physics of Gansu Province, Lanzhou University, Lanzhou, Gansu 730000, China}

\date{\today}

\begin{abstract}

The cover-time problem, i.e., time to visit every site in a system, is one of the key issues of random walks with wide applications in natural, social, and engineered systems. Addressing the full distribution of cover times for random walk on complex structures has been a long-standing challenge and has attracted persistent efforts. Yet, the known results are essentially limited to {\it homogeneous} systems, where different sites are on an equal footing and have identical or close mean first-passage times, such as random walks on a torus. In contrast, realistic random walks are prevailingly {\it heterogeneous} with diversified mean first-passage times. Does a universal distribution still exist? Here, by considering the most general situations, we uncover a generalized rescaling relation for the cover time, exploiting the diversified mean first-passage times that have not been accounted for before. This allows us to concretely establish a universal distribution of the rescaled cover times for heterogeneous random walks, which turns out to be the Gumbel universality class that is ubiquitous for a large family of extreme value statistics. Our analysis is based on the transfer matrix framework, which is generic that besides heterogeneity, it is also robust against biased protocols, directed links, and self-connecting loops. The finding is corroborated with extensive numerical simulations 
of diverse heterogeneous non-compact random walks on both model and realistic topological structures. 
Our new technical ingredient may be exploited for other extreme value or ergodicity problems with nonidentical distributions.
\end{abstract}



\maketitle

\section{\label{sec:level1}Introduction}

Random walk \cite{Weiss1994,Hughes1995,Qian2006RW,Codling2008,Benichou2014,Masuda2017} has been one of the pillars of the probability theory since the 17th century from the analysis of games of chance \cite{Todhunter2006history}, and lays the foundation of the modern theory of stochastic processes and Brownian motions \cite{Einstein1905,Smoluchowski1906A,WangUhlenbeck1945}.
Cover time, the time for a random walker to visit all the sites, is a key quantity that characterizes the efficiency of exhaustive search \cite{Aleliunas1979,kahn1989cover,Nemirovsky1990cover,Aldous1983,Wilf1989}. The cover-time problem, also known as the traverse process \cite{Aleliunas1979,kahn1989cover}, has widespread natural \cite{Viswanathanetal1999,Moreau2011,Peter2017cover,Heuze2013}, social \cite{Feller1968,Aldous1983,Wilf1989,BonehHofri1989}, and engineering \cite{Wang2001PRE,Belardinelli2007,Vergassola2007} applications.
Examples include
rodent animals searching and storing as much food as possible in their confined habitats \cite{Viswanathanetal1999,Moreau2011,Peter2017cover},
the dendritic cells chasing all danger-associated antigens in a constantly changing tissue environment~\cite{Heuze2013},
collecting all the items in the classic coupon collector problem
\cite{Feller1968, Aldous1983,Wilf1989, BonehHofri1989},
the Wang-Landau Monte Carlo algorithm sampling every energy state in calculating the density of states in a rough energy landscape~\cite{Wang2001, Wang2001PRE, Belardinelli2007},
robotic exploration of a complex domain for cleaning or demining and corresponding algorithm design \cite{Vergassola2007}, and information spreading or collecting on large scale Internet, mobile {\it ad hoc} network, peer-to-peer network, and other distributed systems where random walks are more feasible versus topology-driven algorithms due to the dynamical evolution, unknown global structure, limited memory, or otherwise broadcast storm issues \cite{Gkantsidis2004,Santos2005,Avin2008, MianBeraldi2010, Li2012wireless}.

Since the proposal of the cover-time problem~\cite{Aleliunas1979,Aldous1983,kahn1989cover,Wilf1989,Nemirovsky1990cover}, a series of theoretical progresses has been made for standard random walks, where the walker moves to the neighboring sites with equal probabilities.
By bridging the cover time with the longest first-passage time (FPT, the time required to reach a particular site), the cover time can be estimated via the tail of the FPT distribution \cite{Aldous1983}. The average cover time scales distinctively in different dimensions, i.e., $N^2$ for one-dimensional (1D) lattices \cite{Wilf1989,yokoi1990some,Nemirovsky1990cover}, $N(\ln N)^2$ for two-dimensional (2D) lattices with periodic boundary conditions, or 2-torus \cite{Aldous1991cover2D,Peter2017cover}, and $N \ln N$ for 3- or 4-torus \cite{Nemirovsky1990cover}. 


Despite these theoretical progresses, the full distribution of the cover time for complex topologies has been a long-standing challenge and has attracted persistent efforts due to its ultimate importance in characterizing all types of statistics including extreme events \cite{ErdosRenyi1961,Aldous1989book,Belius2013,Chupeau2015cover,Maziya2020,Maier2017}.
Erd\H{o}s and R\'{e}nyi found in 1961 that the cover time for fully connected graphs with self-connecting loops (the coupon collection time) follows the Gumbel distribution \cite{ErdosRenyi1961}, one of the well-known extreme value distributions \cite{castillo2012extreme}.
Later in 1989 Aldous conjectured that for random walks on $d\geq 3$ torus the distribution is also Gumbel \cite{Aldous1989book}, which has been proved by Belius in 2013 \cite{Belius2013}.
Recently, a substantial progress has been achieved for non-compact {\it homogeneous} random walks, where the mean first-passage times (MFPT) $\langle T_k \rangle \equiv \langle T_{k\leftarrow s}\rangle_s$ averaging over the starting sites are narrowly distributed and
can be approximated by a single value $\langle T \rangle \equiv \langle T_k \rangle_k$, i.e., the global mean first-passage time (GMFPT).
In particular, by rescaling the cover time $\tau$ as
$\tilde{\chi}  = \tau/\langle T \rangle - \ln N$ and
employing the Laplace transform, the full distribution for $\tilde{\chi}$ was shown to follow the Gumbel universality class~\cite{Chupeau2015cover}.

It should be noted that the theory in Ref. \cite{Chupeau2015cover} only depends on a single value, i.e., the GMFPT $\langle T \rangle$, of different random walk models. 
This will be sufficient for {\it homogeneous} systems. For example, for random walks on a torus, $\langle T_k \rangle$ for different sites will be exactly the same due to the translational symmetry.
However, for confined or disordered systems, or those with complex topological structures, as illustrated in Fig. 1,
sites at different locations are obviously inequivalent \cite{Dong1999Fluid,Condamin2005, ZoiaCortis2010,HwangKahngPRE2014,Benichou2014,Manzoetal2015,Cheng2018granular,Luo2018NonGaussuan,Giuggioli2020}, even in the thermodynamic limit.
This invalids the basis of the rescaling process and leads to an immediate question that whether there still exists a universal cover-time distribution for such systems, and more commonly, in general
{\it heterogeneous} systems where $\langle T_k \rangle$ can be diverse \cite{HwangKahng2012}.
This is of particular importance as most realistic systems are in fact heterogeneous.


\begin{figure}
\center
\includegraphics[width=\linewidth]{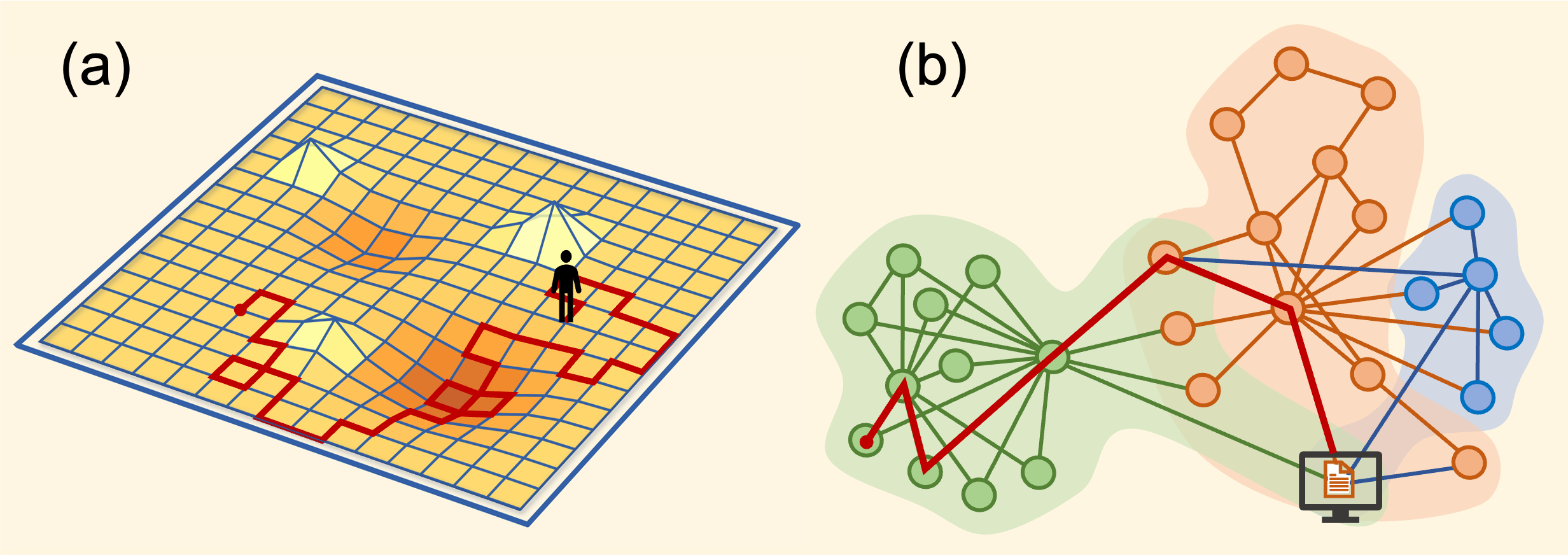}
\caption{{Schematics of heterogeneous random walks.} (a) Heterogeneity due to the confinement boundary and nonuniform landscapes, which result in diversified location-dependent occupation ratios and mean first-passage times. (b) Heterogeneity due to inherent heterogeneous connection structures, e.g., a small subset pruned from Wikipedia where nodes are pages and edges are hyperlinks.
}
\label{fig:3D_FPT}
\end{figure}

In this paper, based on the transfer matrix framework
and employing the longest FPT for the cover time, we derive the Gumbel universality class for rescaled cover-time distributions explicitly in heterogeneous systems under reasonable approximations. The key is the generalized rescaling relation exploiting the complete set of MFPTs $\{\langle T_k\rangle\}$,
which can degenerate to
Ref. \cite{Chupeau2015cover} for homogeneous cases.
Thus our results extend the universality of the Gumbel class to realistic, heterogeneous non-compact random walks.
The finding is corroborated by extensive numerical simulations of 12 diverse random walk models with ideal or realistic heterogeneous structures, standard or biased random walk protocols, and undirected or directed connections. 

The rest of the paper is organized as follows.
In Sec.~\ref{sec:theory}, the distribution function of the rescaled
cover times in heterogeneous systems is derived.
Section~\ref{sec:sim} provides evidence of the universal distribution with extensive simulations.
Conclusion and discussions are provided in Sec.~\ref{sec:conclude}. Detailed derivations are provided in the Appendices.

\section{Theory of universal cover-time distributions for heterogeneous random walks}\label{sec:theory}

For random walks in homogeneous systems, different sites are on an equal footing where the MFPT to all the sites are close to each other and can be approximated by their mean value, i.e., the GMFPT $\langle T \rangle$. In contrast, for random walks in a confined region, the existence of the boundary breaks the symmetry of the sites.
For example, the boundary itself may have attracting effects that the walker crawls along the boundary and reaches sites closer to the boundary with a higher probability.
Therefore, the FPT between a given pair of sites depends not only on their distance, but also on their specific locations. This is even critical for random walks on complex structures such as heterogeneous graphs, where sites with more edges are easier to be reached. As a natural consequence, the time for the walker to visit all the sites, i.e., the cover time, will need to consider the diversity of those FPTs, especially the long FPTs.

Therefore, in the following we shall first present the FPT distribution function based on the transfer matrix framework of Markovian process, which can take the heterogeneity into account through inhomogeneous transition probabilities. Then by counting the longest FPTs, the distributions of the cover times are derived.

\subsection{First-passage time}

The transfer probability that a random walker moves from site $j$ to site $i$ is denoted as $\omega_{i \leftarrow j}$. The case $i=j$ is included to account for the case when the walker has a nonzero probability to stay on the same site at the next time step, which effectively describes the effect of self-connecting loops. Collecting all the elements forms the transfer matrix ${\bf\Omega} =[\omega_{i \leftarrow j}]$. The occupation probability $g_i(t)$ that a walker appears on site $i$ at time $t$ can be obtained recursively by
\begin{equation}
g_i(t)=\sum_j \omega_{i \leftarrow j} g_j(t-1).
\label{eq:basic}
\end{equation}
To obtain the FPT from starting site $s$ to site $k$, a matrix ${\bf D}(k)$ is constructed based on Eq.~(\ref{eq:basic}). The element ${D}_{ij}(k)$ equals $\omega_{i \leftarrow j}$ except the $k$'th column ${D}_{ik}(k)$, which equals 0. When $\omega_{i \leftarrow j}$ in Eq.~(\ref{eq:basic}) is replaced by $D_{ij}(k)$, if the random walker arrives at site $k$, it will be removed from the system in the next time step. Then, the FPT probability $F_{k\leftarrow s}(t)$ from site $s$ to site $k$ at time $t$ is equal to the difference of the probability that the walker is still in the system at time $t$ and that at $t+1$, given that the walker starts from site $s$ in the beginning:
\begin{equation}
F_{k\leftarrow s}(t) =  \Vert {\bf D}(k)^{t} {\bf G}(s) \Vert_1 - \Vert {\bf D}(k)^{t+1} {\bf G}(s) \Vert_1,
\label{eq:fpt}
\end{equation}
where ${\bf G}(s)$ is the initial spatial distribution at $t=0$, $G_i(s)=1$ if $i=s$ and zero otherwise, $\Vert {\bf v} \Vert_1 = \sum_i v_i$ is the L1 norm of vector ${\bf v}$.

In very short time scales, $F_{k\leftarrow s}(t)$ is caused by the direct diffusion process according to the transfer matrix, thus it depends on the specific starting site $s$ and could decrease rapidly versus time if $s$ and $k$ are not far from each other.

For large $t$, the asymptotic behavior of $F_{k\leftarrow s}(t)$ in Eq.~(\ref{eq:fpt}) is determined by the largest eigenvalue $\lambda_k$ ($0<\lambda_k<1$) of the matrix ${\bf D}(k)$, i.e., $F_{k\leftarrow s}(t) \sim \lambda_k^t \sim e^{-t/T_k}$, where $T_k= -1/\ln(\lambda_k)$ is the characteristic FPT to site $k$ in the long time limit.
Therefore, the asymptotic behavior of Eq.~(\ref{eq:fpt}) can be written as
\begin{equation}
F_{k}(t) \simeq \frac{1}{T_k} \exp\left(-\frac{t}{T_k}\right).
\label{eq:fpt2}
\end{equation}
The FPT distribution $F_{k\leftarrow s}(t)$ in the long time limit only depends on $T_k$, but is independent to the starting site $s$.
For random walk on homogeneous systems such as lattices with periodic boundary conditions, all target sites are identical and share the same characteristic FPT. However, for heterogeneous systems such as confined region with nontrivial boundary conditions or other complex structures, the characteristic FPT $T_k$ can be scattered in a broad span, thus the distribution function $F_k(t)$ will be nonidentical and site dependent.
Therefore, the complete set of the diverse characteristic FPTs will be needed to determine the cover-time distribution in heterogeneous system.

In many cases it is impractical to calculate the characteristic FPT $T_k$ based on matrix ${\bf D}(k)$, since the transfer probability $\omega_{i \leftarrow j}$ can not be completely obtained in general. Alternatively, since $T_k=\int tF_{k}(t) {\rm d}t$ is also the mean FPT to site $k$,
$T_k$ can be approximated by the average of the FPT $T_{k \leftarrow s}$ from all possible starting site $s$ to site $k$, i.e., $T_k \approx \langle T_k \rangle \equiv \langle T_{k \leftarrow s}\rangle_s$ (see Appendix \ref{sec:Tk} for an exemplary numerical verification).

\subsection{Full cover time}

For a trajectory which fully covers the system, the random walker will first successively pass $N-1$ sites in a specific order. Then the time arriving at the last unvisited site is the full cover time. Meanwhile, this time is also the FPT to this ending site. Therefore, the cover time can be regarded as the longest FPT in a single cover process \cite{Aldous1983}.
To be specific, the probability that the cover time equals $\tau$ can be estimated as
\begin{equation}
\begin{aligned}
P(\tau) = \frac{1}{N}\sum_{k,s}^{k\neq s} F_k(\tau) \left[ \prod_{i}^{i\notin \{k, s\}} \sum_{t}^{\tau-1} F_i(t) \right].
\end{aligned}
\label{eq:cover}
\end{equation}
The formula in the square brackets denotes the probability that all sites except the starting site $s$ and the ending site $k$ have been visited during $\tau-1$ steps, and $F_k(\tau)$ is the probability that the last unvisited site $k$ is firstly visited at $\tau$, which completes the covering process. The factor $1/N$ is for the average over all the starting sites.
Equation (\ref{eq:cover}) neglects any structural correlation of the background system, which should be (at least weakly) satisfied for non-compact random walks \cite{Gennes1982compact}, i.e., the
diffusion dimension $d_w$ of the random walk (defined by $\langle |{\bf r}|^2 \rangle \sim t^{2/d_w}$) is smaller than the dimension of the background space $d$.
In this case, the number of sites visited by the walker $|{\bf r}|^{d_w}$ will be negligible compared to all the sites within $|{\bf r}|$, which is $|{\bf r}|^d$. Thus the walker needs to traverse the region many times in order to cover all the sites. This means that the effective structural correlation will be significantly reduced as the information of the initial position is completely lost when the walker comes back again.

Replacing $T_k$ by $\langle T_k \rangle$, and considering $\tau - 1 \cong \tau$ for $\tau \gg 1$, the summation in the square bracket yields
\begin{equation}
\sum_{t=1}^{\tau} F_k(t) \cong 1 - \exp\left(-\frac{\tau}{\langle T_k \rangle}\right).
\label{eq:mid0}
\end{equation}
For $\tau$ being large, higher order terms in the production in the square bracket can be neglected, leading to
\begin{equation}
\prod_{i} \sum_{t}^{\tau} F_i(t)
\cong \exp\left( - \sum_{i=1}^{N} \exp\left(- \frac{\tau}{\langle T_i \rangle}\right) \right),
\label{eq:A:mid2}
\end{equation}
where the condition $i\notin \{k, s\}$ has also been relaxed. Define a rescaled cover time $\chi$ from the original cover time $\tau$ by
\begin{equation}
\chi= -\ln \sum_i e^{- \frac{\tau}{\langle T_i \rangle} },
\label{eq:xstar}
\end{equation}
where the summation can be replaced by the integration if the distribution of $\langle T_i \rangle$ is known, i.e., $\sum_i e^{- \frac{\tau}{\langle T_i \rangle} } = N \int e^{- \frac{\tau}{\langle T_i \rangle} }P({\langle T_i \rangle}){\rm d}{\langle T_i \rangle}$.
Together with Eq. (\ref{eq:fpt2}), it is straightforward to show that $\sum_{k}F_k(\tau)=\exp(-\chi)\mathrm{d}\chi/\mathrm{d}\tau$. Thus one has
\begin{equation}
P(\tau){\rm d}\tau \cong  \exp[-\exp(-\chi)] \exp(-\chi) {\rm d}\chi,
\label{eq:A:mid4}
\end{equation}
yielding a universal distribution function for the rescaled cover time $\chi$ that is independent to any of the system details:
\begin{equation}
P(\chi) = \exp(-\chi-\exp(-\chi)).
\label{eq:cover2}
\end{equation}
The detailed derivations are shown in Appendix~\ref{sec:fullcover}.

Equation~(\ref{eq:cover2}) is the Gumbel distribution which is one of the three well-known extreme value distributions~\cite{beirlant2006statistics}, and shares the same form as in the theory for homogeneous cases \cite{Chupeau2015cover}. The main difference is the rescaling functional relation between $\chi$ and $\tau$, e.g., $\tilde{\chi} = \tau/\langle T \rangle - \ln N$ for homogeneous cases, and Eq.~(\ref{eq:xstar}) for the heterogeneous cases, which now requires the complete set of MFPTs $\{\langle T_k \rangle\}$ or their distribution function. Equation~(\ref{eq:xstar})
ensures a monotonous relation between $\chi$ and $\tau$.
For homogeneous cases, all sites share a single identical characteristic MFPT, which is then the GMFPT $\langle T \rangle$, and Eq. (\ref{eq:xstar}) degenerates to $\tilde{\chi} = \tau/\langle T \rangle - \ln N$.

Note that in the derivation of Eq. (\ref{eq:cover2}), it is assumed that the cover time is large compared to the MFPT $\langle T_k\rangle$. This condition is typically satisfied automatically, as the cover time $\tau$ ending at site $k$ is approximately the longest FPT to $k$.
Therefore, the large FPT dominates in the covering process, justifying the use of Eq. (\ref{eq:fpt2}). Even for short $\tau$, e.g., negative $\chi$, it is usually still larger than $\langle T_k\rangle$, thus Eq. (\ref{eq:cover2}) works well.

\subsection{Partial cover time}

For a random walk that the walker only needs to visit a fraction of the system, it is obvious that the time cost depends on how to choose the unvisited sites. There are two related processes, one is random cover, and the other is partial cover. In the random cover problem, the $m$ sites that are not counted (but can be visited) are chosen randomly from the total $N$ sites in the system before the cover process.
For $m \ll N$, the randomly chosen sites are probably already visited during the random cover process to visit all the other sites, therefore it will have almost the same cover-time distribution with the full cover process. In the partial cover process, it stops when there are $m$ sites left. If the $m$ sites are distributed randomly in the system, for the full cover process the walker needs to wander over the system again and again until it finds all of them, which will in general take a much longer time, at least in the order of their respective MFPTs.
Therefore, the partial cover times will have a nontrivial distinctive distribution other than that for random or full cover processes.

Although significantly different from the full cover process, for partial cover there still exists a universal distribution for non-compact random walks in heterogeneous systems. In particular, the partial cover-time distribution can be expressed as
\begin{equation}
P_m(\tau) = \sum_{\{I_m\}}
P(\tau | \notin \{I_m\}) \, Q(\tau , \{I_m\}),
\label{eq:partial}
\end{equation}
where $\{I_m\}$ is a specific set of $m$ sites, and we assume $m\ll N$. The first term $P(\tau|\notin \{I_m\})$ denotes the probability that the walker visits the $N-m$ other sites during time $\tau$ regardless of the sites in the set $\{I_m\}$ being visited or not. This follows the random cover-time distribution, which can be approximated by the full cover-time distribution.

The second term $Q(\tau, \{I_m\})$ is the probability that the sites in set $\{I_m\}$ are not visited yet at time $\tau$.
If the unvisited sites $\{I_m\}$ are independent, i.e., the information of the sites that are already visited in $\{I_m\}$ does not change the probability that any of the rest sites in $\{I_m\}$ to be visited, the second term $Q$ can be simplified as
\begin{equation*}
Q(\tau , \{I_m\}) = \prod_i^{\{I_m\}} \left(1-\sum_t^{\tau} F_i(t) \right) = \prod_{i}^{\{I_m\}} \exp\left(-\frac{\tau}{\langle T_i \rangle}\right),
\end{equation*}
where the second equality takes use of Eq. (\ref{eq:mid0}).

After enumerating all possible configurations of $\{I_m\}$ by the summation $\sum_{\{I_m\}}$, the partial cover-time distribution $P_m(\tau)$ can then be obtained in the form of Gumbel class (Appendix~\ref{secA:partial})
\begin{equation}
P_{m}(\chi) = \frac{1}{m!}\exp\left(-(m+1)\chi-e^{-\chi}\right),
\label{eq:partial2}
\end{equation}
where $\chi$ is the rescaled cover time and is defined by Eq.~(\ref{eq:xstar}).  According to Eq.~(\ref{eq:partial2}), the most probable rescaled partial cover time $\chi^*$ satisfies $\exp(-\chi^*)=1+m$, or $\chi^*=-\ln (1+m)$. Thus $\chi^*$ is characteristically distinct for different $m$. Due to the monotonous relation between $\chi$ and $\tau$, this indicates that the corresponding characteristic time $\tau^*$ to visit $N$ ($m=0$) and $N-1$ ($m=1$) sites are discontinuous, which is consistent with the results in \cite{Coutinho1994pcover} for full cover and partial cover with $m=1$ in 2D lattices.

For correlated unvisited sites, since $P(\tau|\notin \{I_m\})$ is approximately the random (or full) cover-time distribution, the main issue caused by the correlation is the simplification of the $Q$ term.
When $N \gg m$, the correlation of the unvisited sites can be weak, and it will lead to only a small correction. As a consequence, $m$ will need to be replaced with an effective, typically smaller, $m^*$, and the rescaled partial cover-time distribution has the same form as Eq. (\ref{eq:partial2}).
Indeed, it has been found that for various heterogeneous random walk models, the rescaled partial cover-time distributions of different systems can collapse to an identical distribution function that agrees perfectly with Eq.~(\ref{eq:partial2}).
However, there are also cases that correlation of unvisited sites could be so strong that can not be ignored even in the large $N$ limit, especially for low dimensional random walks, such as 2D persistent random walks and random walk on 3D lattices with reflective boundaries.

\subsection{Parallel search}

Parallel search is an important practical issue. We have examined the MFPT and the rescaled cover-time distribution when there are $n$ ($\ll N$) independent walkers in heterogeneous systems. In particular, in the long time limit, the MFPT to site $k$ follows $n$ independent and identical distributions of Eq. \ref{eq:fpt2}, which is then $F_{k}^{(n)}(t) \simeq \frac{n}{T_k} e^{-\frac{nt}{T_k}}$. Thus the new characteristic FPT $T'_k$ (also the MFPT $\langle T'_k\rangle$) becomes $1/n$ of that for one walker. With the new set of $\{\langle T'_k\rangle\}$, the rescaled cover time can be obtained from Eq. \ref{eq:xstar} to yield the universal distributions Eqs. \ref{eq:cover2} and \ref{eq:partial2}. 
Thus when there are $n$ independent parallel walkers in a heterogeneous system, the cover time will be $1/n$ as that for one walker.

It should be noted that our approach to derive the full and partial cover-time distribution is based on the general framework of transfer matrix, whose elements can be assigned arbitrarily given the normalization condition that the summation of the elements for each column is one. It is not necessary that the walker moves to the neighbors with equal probability as in the standard random walk models. Indeed, the results are also valid for biased walking protocols or directed connections, as will be demonstrated in the simulation part. In addition, the walker may have a nonzero probability to stay on the same site for the next step, accounting for the self-connecting loop effects.
These considerations enable our theory to be applicable in broad realistic circumstances.

\section{Numerical verification}\label{sec:sim}

In this section, the universality of the rescaled cover-time distribution will be corroborated with different types of random walks on various heterogeneous systems, for both full cover [Eqs.~(\ref{eq:cover2}) and (\ref{eq:xstar})] and partial cover [Eqs.~(\ref{eq:partial2}) and (\ref{eq:xstar})] processes. 
Here we provide extensive numerical simulation results for 12 cases with different random walk models and topological structures, including 3 realistic networks, 
which agree with the full cover theory well. For partial cover, most of the simulation cases (10) agree, but deviation occurs for low dimensional cases 
where the correlation of the unvisited sites is not negligible.


\begin{figure}
\includegraphics[width=1\linewidth]{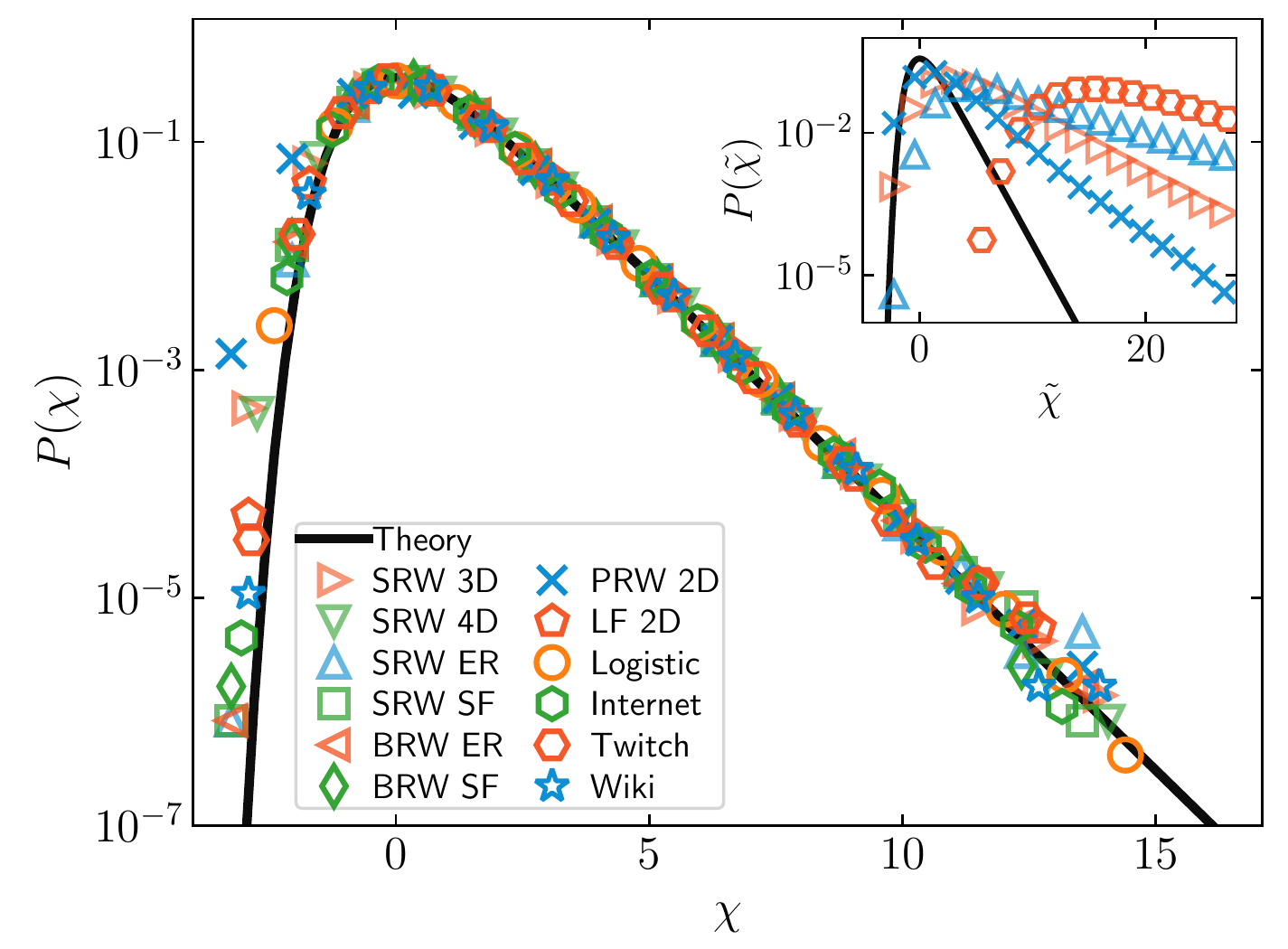}
\caption{Rescaled full cover-time distributions of the 12 random walk cases. The key parameters are as follows. SRW 3D: $N=343$; SRW 4D: $N=625$; SRW ER: $N=1000$, $\langle K\rangle =8$; SRW SF: $N=2000$, $\langle K\rangle =4$; BRW ER: $N=1000$, $\langle K\rangle =8$; BRW SF: $N=2000$, $\langle K\rangle =4$; PRW 2D: $N=2304$, $l_c=24$; LF 2D: $N=1600$; Logistic: $N=1000$; Internet: $N=11174$; Twitch: $N=7126$; Wiki: $N=3623$.
The cover-time distribution for each case is evaluated from sampling over 1 million rounds.
The symbols are the data rescaled with Eq.~(\ref{eq:xstar}), and the thick black curve is Eq.~(\ref{eq:cover2}).
Inset: a few representative cases where the cover time is rescaled based only on the GMFPT as $\tilde{\chi} = \tau/\langle T \rangle - \ln N$. The thick black curve is the corresponding prediction.
}
\label{fig:fig2}
\end{figure}

The 12 random walk cases are: 1) standard random walk on 3D lattice with reflective boundaries (SRW 3D); 2) standard random walk on 4D lattice with reflective boundaries (SRW 4D); 3) standard random walk on Erd\H{o}s-R\'{e}nyi graph (SRW ER); 4) standard random walk on scale-free graphs (SRW SF);
5) biased random walk on Erd\H{o}s-R\'{e}nyi graph (BRW ER); 6) biased random walk on scale-free graphs (BRW SF);
7) persistent random walk on 2D lattice with reflective boundaries (PRW 2D); 8) the L{\'e}vy flight on 2D lattice with sticking boundaries (LF 2D); 9) chaotic motion as random walks (Logistic); 10) standard random walk on the Internet at the autonomous system level (Internet); 11) standard random walk on the Twitch social network (Twitch); and 12) standard random walk on the directed sub-graph containing the page ``Random walk'' of Wikipedia (Wiki). The topological structures of the last three cases come from the real-world complex systems.
These models are chosen due to their simplicity, representativeness, and significance in realistic applications.

In our simulation, a site $s$ is chosen randomly from a total of $N$ sites with probability $1/N$ as the starting site, then the random walk begins, until it covers the whole system, yielding the cover time $\tau$. In the meantime, the partial cover time with arbitrary $m$ and one ensemble of the FPT from the starting site $s$ to all the other sites $T_{k \leftarrow s}$ are also obtained. For each case, an overall of 1 million rounds is carried out to obtain reliable statistics. For each $\tau$, the corresponding $\chi$ is derived according to Eq.~(\ref{eq:xstar}) based on the numerically obtained $\{\langle T_{k}\rangle \}$. Then the distribution of the one million $\chi$ values is obtained by direct counting, which is plotted in Fig.~\ref{fig:fig2}. The different symbols are for different cases.
They are all in good agreement with the universal distribution Eq.~(\ref{eq:cover2}) as plotted by the thick black curve.

All the above 12 cases are heterogeneous. Due to the diversified value of the MFPT $\langle T_{k}\rangle$, the GMFPT $\langle T\rangle$ is insufficient to rescale the cover time by $\tilde{\chi} = \tau/\langle T \rangle - \ln N$, which leads to substantial deviations, as shown in the inset of Fig.~\ref{fig:fig2} for a few representative cases.

A brief description of the models is as follows. Detailed parameters may be listed in the caption of the corresponding figures.

(1) Standard random walk on 3D lattice in a cubic domain with reflective boundary conditions (SRW 3D), i.e., if the walker moves outside the boundary, it will be reflected back to the cubic domain. As a result, the MFPT to the target site is location dependent and mainly depends on the distance to the boundary~\cite{Condamin2005,Meyer2011FPT, Giuggioli2020}.
For a given finite domain, the transfer matrix ${\bf \Omega}$ can be written down explicitly \cite{Ref3D}.

(2) Standard random walk on 4D lattice in a hypercubic domain with reflective boundary conditions (SRW 4D). The transfer matrix ${\bf \Omega}$ can be obtained similarly.

(3) Standard random walk on Erd\H{o}s-R\'{e}nyi graph \cite{erdos59a} (SRW ER).
The ER graph with size $N$ is generated by connecting each pair of nodes with probability $p$. A graph is connected if there is always a path passing through sites and edges to connect any given pair of sites.
In our simulation, only the connected graphs are considered. The degree $K$ of a site is the number of its edges. The degree distribution for ER graph is Poisson: $P(K) = \frac{\langle K \rangle^K}{K!} \exp\left(-\langle K \rangle \right)$, where $\langle K \rangle \approx Np$ for large $N$ is the average degree, which is also the variance.
Thus when $\langle K \rangle$ is large, the relative standard deviation with respect to the mean, i.e., $1/\sqrt{\langle K \rangle}$, is small, and the system is approximately homogeneous. However, when $\langle K \rangle$ is not so large, the degrees and consequently the MFPTs can be broadly distributed, breaking the homogeneous assumption.

For standard random walk on a graph, starting from any site with degree $K$, the walker moves to any of its neighboring sites with equal probability $1/K$.

(4) Standard random walk on scale-free graphs (SRW SF) with power law degree distribution \cite{BAnetwork1999}, i.e.
${P}(K)\sim K^{-\gamma}$ and $\gamma=3$ in our simulation, $\langle K \rangle = \int KP(K){\rm d}K$ is the average degree.
This system is inherently heterogeneous, even in the thermodynamic limit.

(5) Biased random walk on Erd\H{o}s-R\'{e}nyi graph (BRW ER).
Unlike the standard random walk that the walker moves to one of its neighbors with equal probability, for biased random walk, the walker at site $s$ with $K_s$ neighbors moves to a neighboring site $i$ with the probability $K_i^{-\alpha}/\sum_{j=1}^{K_s} K_j^{-\alpha}$, where the summation is over the site $i$'s neighbors. When $\alpha = 0$, it returns to the standard random walk. We take $\alpha = 1$ for this case.

(6) Biased random walk on scale-free graphs (BRW SF), $\alpha = 0.18$.

\begin{figure}
\includegraphics[width=\linewidth]{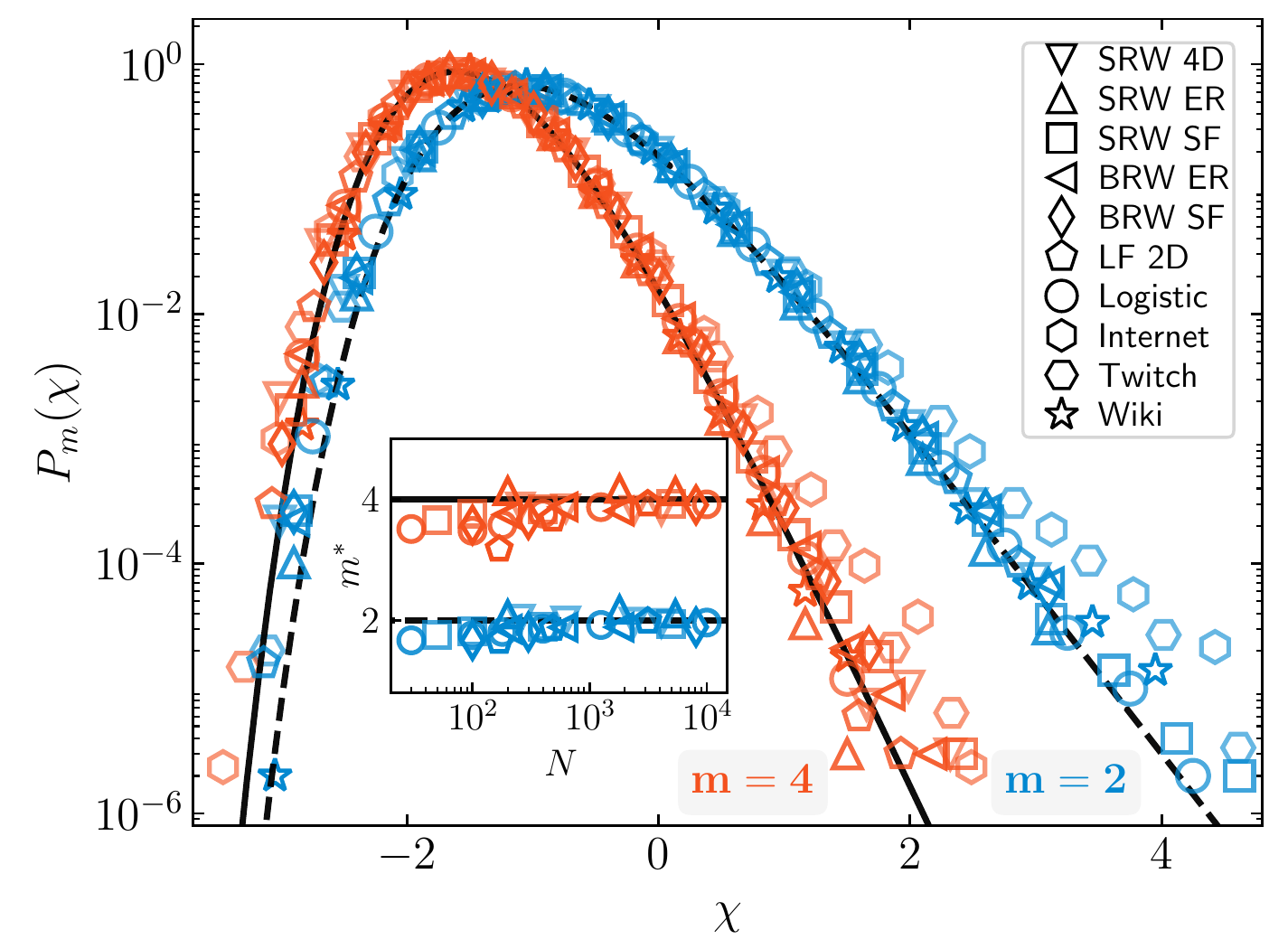}
\caption{Rescaled partial cover-time distributions of 10 heterogeneous random walk models as labeled in the figure. SRW 4D: $N=4096$; SRW ER: $N=5400$; SRW SF: $N=5000$; BRW ER: $N=5400$; BRW SF: $N=2000$; LF 2D: $N=6400$; Logistic map: $N=10000$; Internet: $N=11174$; Twitch: $N=7126$; and Wiki: $N=3623$. Each distribution is obtained from 1 million samples. The symbols are the data, and the curves are the theory Eq.~(\ref{eq:partial2}) for $m=2$ (dashed) and $m=4$ (solid). The inset shows the effective $m^*$ vs size $N$ of the 7 model systems where the size can be changed.
}
\label{fig:fig3}
\end{figure}

(7) The persistent random walk \cite{Tejedor2012,Izzetetal2020} on 2D lattices in a square domain of side length $L$ with reflective boundary conditions (PRW 2D). The walker keeps its previous moving direction with probability $p_r$, and moves to any of the other three directions with probability $(1-p_r)/3$. The mean persistent length is given by $l_c = 1/(1-p_r)$ \cite{Tejedor2012}. When it is much larger than one, the random walk is non-compact. However, when $l_c$ is small, especially when $l_c \sim 1.3$, the persistent random walk degenerates to conventional random walk, which becomes compact. In addition, the persistent random walk is not Markovian, i.e., it not only depends on the current position, but also depends on its previous step, therefore the transfer matrix framework is valid only approximately, leading to deviations from the theory, especially for the short cover times (negative $\chi$).

(8) The L{\'e}vy flight \cite{Shlesinger1989,Shlesinger2009,Moreau2011, Zaburdaev2015,Shlesinger2021} on 2D lattices in a
square domain of side length $L$ with sticking boundaries (LF 2D). The walker flies in one of the four directions ($\pm x$,$\pm y$) to a site of distance $l$ with probability
$P(l) = \frac{1}{\pi} \int_{-\infty}^{\infty} \exp\left(-l_0^a k^a\right) \exp(i k l){\rm d}k$~\cite{Chupeau2015cover},
where $l_0 = L/20$ and $a=3/2$ are constants. If the walker flies out of the region at one time step, it will be stuck at the boundary along its flying route, and continues the flight at the next step.

(9) Chaotic trajectories as a biased and directed random walk (Logistic), which is biased and directed due to the dynamical structures.
A feature for chaotic dynamical systems is that an initial perturbation $\delta(0)$ will be enlarged exponentially in time, i.e., $\delta(t)=\delta(0) \exp(h_1 t)$, where the largest Lyapunov exponent $h_1$ is positive, and $1/h_1$ is the Lyapunov time. The predictability of the chaotic trajectory will vanish after evolving a few Lyapunov times~\cite{ott2018,Lai2020}. Therefore, in time scales that are much larger than the Lyapunov time, the chaotic trajectory can be regarded as a stochastic process. An interesting question is that whether the time for the chaotic trajectory to cover the coarse-grained phase space (the coarse-grained ergodic time) also obeys the universal cover-time distribution.
Here we take the fully chaotic logistic map as an example, which is
$y_{n+1} = 4 y_n (1-y_n)$.
The system is ergodic and the entire phase space can be filled if the evolution time is long enough.
The region $[0,1]$ of the map is divided into $N$ mesh grids. Since the grids have a non-negligible finite size $1/N$, although the original dynamics is deterministic, due to the chaotic nature, the trajectory on the scale of the grids can be highly stochastic.
The occupation ratio is proportional to the natural measure of the system and is highly non-uniform for the logistic map \cite{ott2002chaos}.
This suggests that the MFPT $\langle T_{k} \rangle$ should be also diverse and depends on the position of target grid $k$.

(10) Standard random walk on the Internet at the level of autonomous systems (Internet). This undirected connection structure is a snapshot in 2001 which contains 11174 nodes and 23409 edges \cite{latora2017complex}.

(11) Standard random walk on a subset of the Twitch social network (Twitch). A node is a Twitch user who streams in English, and an edge between two users represents their mutual friendship. This social network contains 7126 users and 35324 edges \cite{Twitch}.

(12) Standard random walk on a directed sub-graph of Wikipedia (Wiki). The sub-graph is pruned from the hyperlink network of Wikipedia, where the original data set is provided in Refs. \cite{Wikipedia1,Wikipedia2}. The node is the page in Wikipedia, and a directed edge represents a hyperlink from one page to another. The original data set has about four million nodes and one hundred million links, which is far beyond our computation power. Therefore, we extract a smaller cluster in which any page can be reached by following at most two steps from the page ``Random walk".  This cluster is strongly connected, that for each pair of nodes ($i,j$), there always exists a directed path in the cluster going from $i$ to $j$.
This sub-graph contains 3623 nodes and 75929 directed edges. The walker moves randomly along the directed edge, which simulates a wanderer exploring the Wikipedia. The probability that the walker moves to a neighboring node (page) is $1/K_o$, where $K_o$ is the out-degree of the current node.

The rescaled cover-time distributions for all the 12 cases are plotted in Fig.~\ref{fig:fig2}. Despite the diversity of the random walk models, the background topology, the biased protocols, and the directed links, the simulation results all fall onto the theoretical curve well, especially for positive $\chi$ values.
As a comparison, the distributions of the rescaled cover time based only on the GMFPT ($\tilde{\chi} = \tau/\langle T \rangle - \ln N$) for a few representative cases, e.g., standard random walks in confined 3D lattices, ER graph, Twitch, and persistent random walks on confined 2D lattices, are shown in the inset of Fig.~\ref{fig:fig2}. They deviate from the prediction significantly.
This illustrates that for heterogeneous systems, rescaling of the cover time only using the GMFPT is insufficient, and the complete set of MFPTs is needed to properly rescale the cover time.

\begin{figure}
\includegraphics[width=\linewidth]{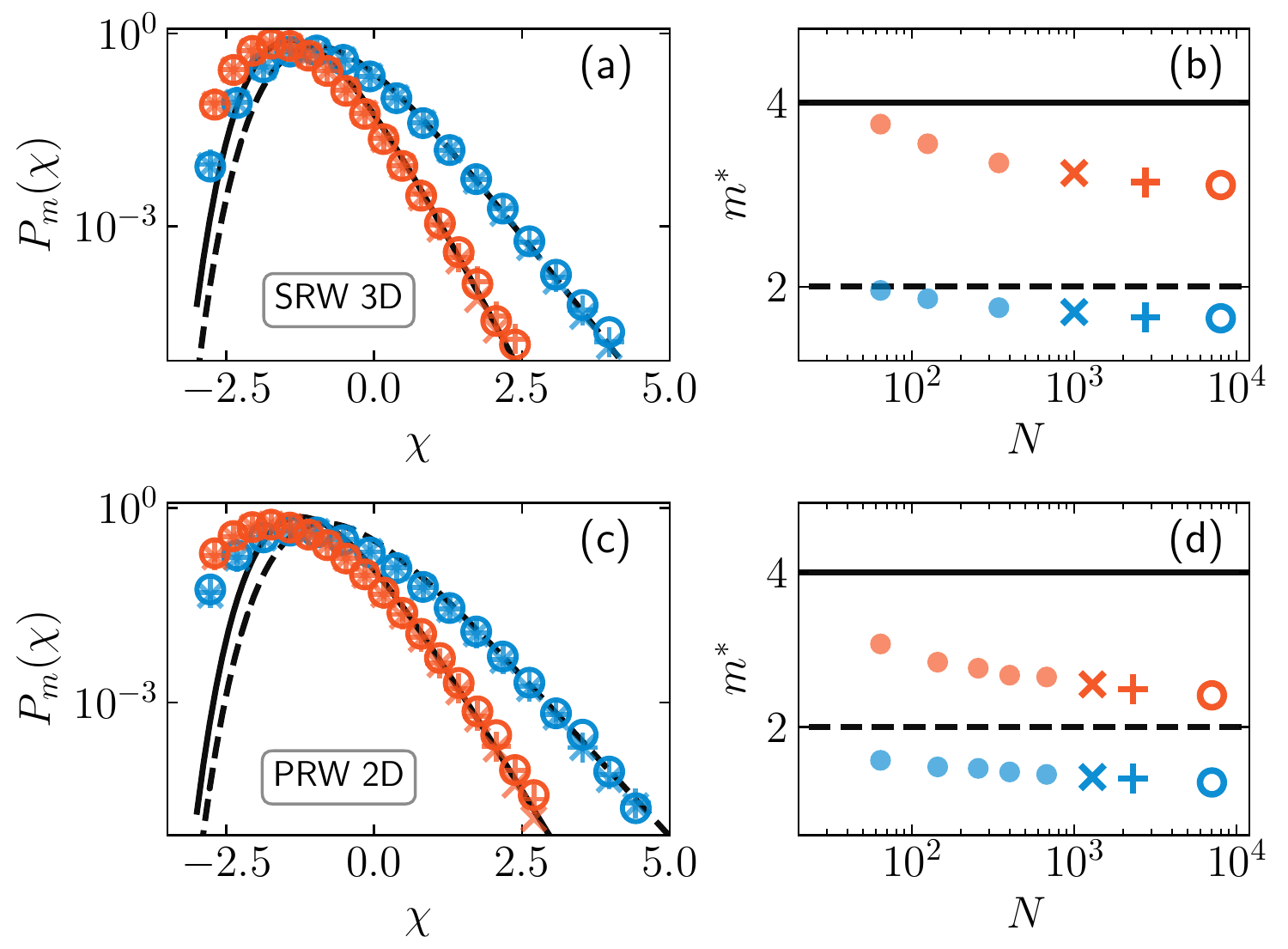}
\caption{(a) Rescaled partial cover-time distribution of SRW 3D. $N=1000, 2744, 8000$ for crosses, pluses, and circles, i.e., the three data points with the largest size in (b). The dashed curve is Eq.~(\ref{eq:partial2}) with $m^* = 1.7$ for $m=2$, and the solid curve is $m^*=3.1$ for $m=4$. (b) The effective $m^*$ vs size $N$ for SRW 3D.
(c) Rescaled partial cover-time distribution of PRW 2D. $N=1296, 2304, 7056$ for crosses, pluses, and circles. The dashed curve is Eq.~(\ref{eq:partial2}) with $m^* = 1.3$ for $m=2$, and the solid curve is $m^*=2.5$ for $m=4$. (d) The effective $m^*$ vs size $N$ for PRW 2D. The characteristic persistent length is $l_c=L/2$.
}
\label{fig:fig4}
\end{figure}

For partial cover time, the 12 heterogeneous random walk models can be classified into two groups. The second group consists of 2 cases: the standard random walk on 3D lattices and persistent random walk on 2D lattices with reflective boundaries. The rest 10 cases are the first group, where the correlation between the unvisited sites is small.
The rescaled partial cover-time distributions of these 10 systems are shown in Fig.~\ref{fig:fig3}, which agree well with the theoretical prediction Eq.~(\ref{eq:partial2}). The correlation between unvisited sites can be treated by a modified
value of $m$, i.e., $m^*$, in Eq. (\ref{eq:partial2}), which can be obtained by fitting the data to the curve Eq.~(\ref{eq:partial2}). The difference between $m^*$ and $m$ then characterizes the extent of correlation.
The inset of Fig.~\ref{fig:fig3} shows the dependence of $m^*$ on the system size. It can be seen that the overall values of $m^*$ are close to $m$, and as the system size increases, $m^*$ converges to $m$. This is consistent with the expectation that as the size of the system increases, the correlation between the unvisited sites decreases.

For the second group, they are both low dimensional cases, where correlation between unvisited sites could be non-negligible and leads to a larger difference between $m^*$ and $m$. The rescaled cover time and the dependence of $m^*$ versus system size for these two cases are shown in Fig.~\ref{fig:fig4}. With the effective $m^*$, the rescaled partial cover-time distribution again follows Eq.~(\ref{eq:partial2}) reasonably well, especially in the large cover time limit [Fig.~\ref{fig:fig4}(a,c)]. However, in contrast to the first group, here $m^*$ is close to $m$ only for small system sizes. As the system size increases, the difference between $m^*$ and $m$ does not vanish, instead it becomes larger and $m^*$ approaches to a constant but smaller value than $m$, as shown in Fig.~\ref{fig:fig4}(b,d). Thus this effect could even persist in the thermodynamic limit. Furthermore, as shown in Fig.~\ref{fig:fig4}(c), for each $m$, the cover time for systems with different size can still be rescaled to follow the same distribution, as the data points collapse on each other in almost all $\chi$ ranges, although the deviation from Eq. (\ref{eq:partial2}) at negative $\chi$ is apparent. This indicates that although the ``real'' rescaled distribution function for partial cover time might be different from the theoretical expectation, the scaling relation Eq. (\ref{eq:xstar}) still holds.

\begin{figure}
\center
\includegraphics[width=\linewidth]{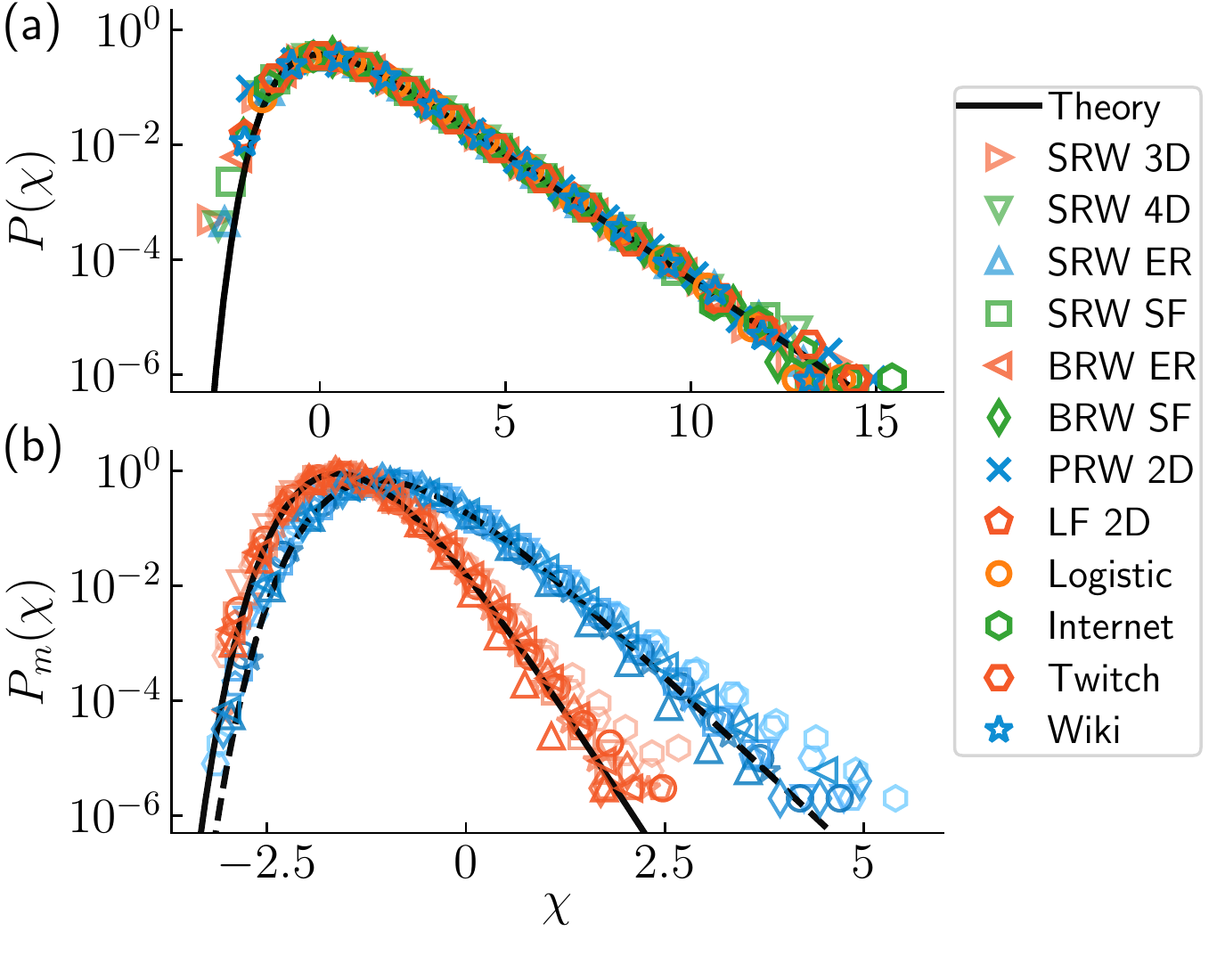}
\caption{The cover-time distribution when there are 5 independent walkers for the simulation models. (a) The full cover-time distribution of the 12 systems in Fig. \ref{fig:fig2}. (b) The partial cover-time distribution of the 10 systems in Fig. \ref{fig:fig3} with dashed and solid lines for $m=2$ and $m=4$, respectively.}
\label{fig-parallel}
\endcenter
\end{figure}

For parallel search, when there are $n$ independent walkers, the new MFPT $\langle T'_k\rangle$ becomes $1/n$ of that for one walker. With the new set of $\{\langle T'_k\rangle\}$, the rescaled cover time can be obtained from Eq. (\ref{eq:xstar}) to yield the universal full and partial cover-time distributions Eqs. (\ref{eq:cover2}) and (\ref{eq:partial2}). Here we provide numerical evidence for the validity of the above reasoning. Figure \ref{fig-parallel} shows the results of both the full and the partial cover-time distributions when there are 5 independent walkers in the simulation systems. It is clear that the distributions from the simulation agree with the theoretical curves well.

\section{\label{sec:conclude} Conclusion and discussions}

The full distribution of cover time for random walks on complex topological structures has been a long-standing issue with broad potential applications.
It has been found widely and robustly in various {\it homogeneous} non-compact random walk models that the cover times can be rescaled to follow the Gumbel distribution \cite{ErdosRenyi1961, Aldous1989book,Belius2013,Chupeau2015cover,Maziya2020}. These results assume that the MFPTs are identical or close to each other. However, in realistic systems such as confined domains or those with inhomogeneous components \cite{Dong1999Fluid,Condamin2005, ZoiaCortis2010, Chevalier2010kinetics, HwangKahngPRE2014, Benichou2014, Manzoetal2015, GodecMetzler2015, KimLee2016, DentzLester2016,Cheng2018granular,Luo2018NonGaussuan, Giuggioli2020, Liu2020walk}, heterogeneity is prevailing, leading to broad spans of the MFPT values \cite{HwangKahng2012, GrebenkovPRE2018}.

In this work, based on the transfer matrix framework,
we have established the universal distribution in {\it heterogeneous} non-compact random walks for both full and partial cover processes. Our results show that the rescaled cover times fall again onto the Gumbel universality class.
Thus we step forward and ground the Gumbel universality of the cover-time distribution on a much broader field of random walks that are prevalent in realistic circumstances.
The key is the rescaling relation, Eq. (\ref{eq:xstar}), where the complete set of the MFPTs $\{\langle T_k \rangle\}$ or their distribution is needed to account for their diversified values.
The results have been corroborated by extensive numerical simulations on 12 random walk models with various background topologies. 
Thus our approach is valid for generalized random walks that can be either standard or biased, undirected or directed, and can have self-connecting loops.

Since heterogeneity is prevailing in realistic random walk situations, e.g., either due to the spatial confinement on the 2D terrestrial surface or in 3D by the cell membrane, building structure, etc., or due to the inherent heterogeneity in the social and engineered systems, our results are expected to have broad applications. For example, our results can be used as a guideline in the design of searching or demining algorithms of robots in complicated environments, or control of the exhaustive information collecting or broadcasting in heterogeneous cyberspace or constantly evolving distributed sensing and communicating systems.

As our theory solves the distribution function of the extreme value problems with nonidentical distributions, it may be exploited in investigating other extreme value problems such as extreme climate events, robustness of engineered systems, etc. \cite{castillo2012extreme,Katz2014climate, Eliazar2019,Geng2021Extreme} with heterogeneous characteristic spatial or time scales. In addition,
the cover process of chaotic trajectories in discretized phase space is effectively the ergodicity issue. Thus our treatment can also be used to estimate the ergodic time of chaotic dynamical systems for a given discretization (or observation resolution) of the phase space with nonuniform natural measures.

\begin{acknowledgments}
This work is supported by NSFC under Grants
No.~12175090, No.~11775101, No.~11905087, No. 12135003, and No.~12047501, and by China Postdoctoral Science Foundation 2019M660810.
\end{acknowledgments}


\appendix

\section{\label{sec:Tk} Approximation of $T_k$ by $\langle T_k \rangle$}

\begin{figure}[h]
\includegraphics[width=\linewidth]{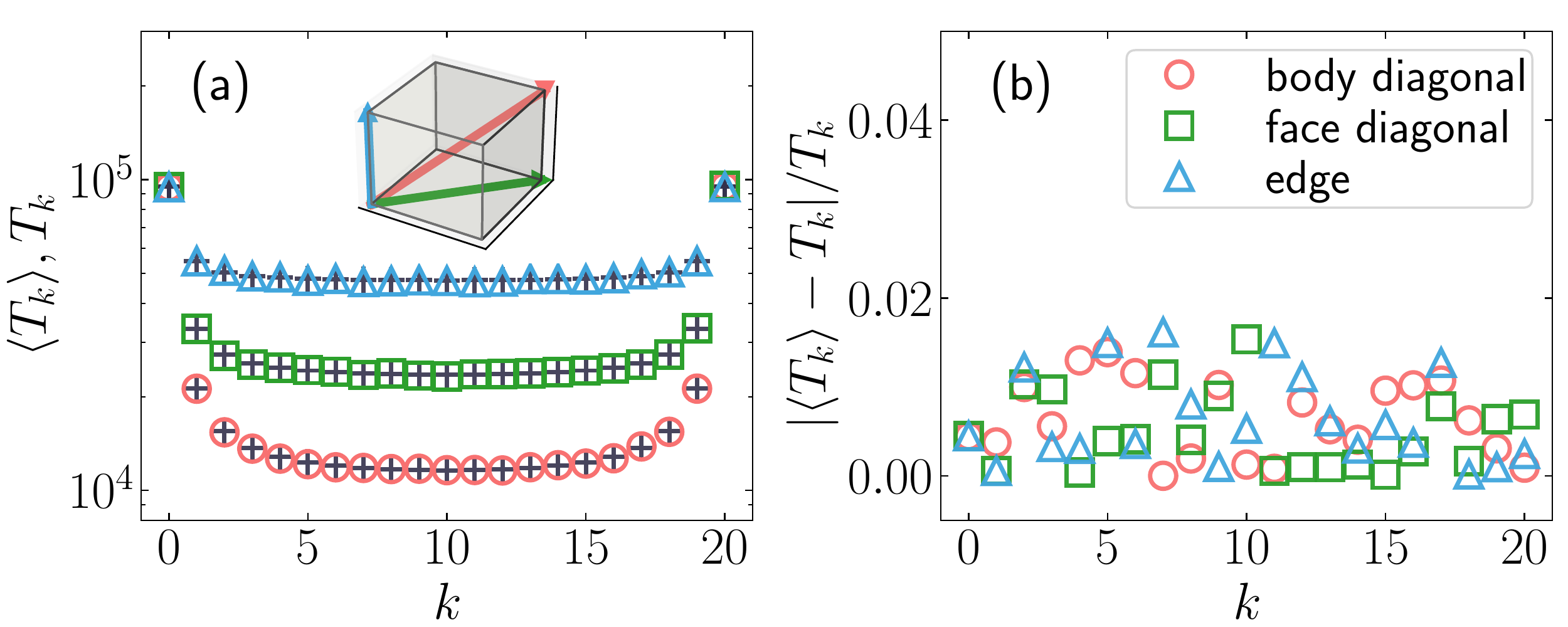}
\caption{{(a)} For standard random walk on a cube of 3D lattice with reflective boundary conditions, the MFPT $\langle T_k \rangle$ and the characteristic FPT $T_k$ for sites along the body diagonal, face diagonal, and the edge of the cube. Each data point of $\langle T_k \rangle$ (empty triangles, squares, and circles) is the average over 20000 simulations, $T_k$ (the plus symbols) is calculated from ${\bf D}_k$ according to $T_k = -1/\ln(\lambda_k)$, where $\lambda_k$ is the largest eigenvalue of ${\bf D}_k$. {(b)} The relative difference between $\langle T_k \rangle$ and $T_k$. }
\label{fig:Tks}
\end{figure}

Figure~\ref{fig:Tks} compares the MFPT $\langle T_k \rangle$ and the characteristic FPT $T_k$ of standard random walk for representative sites in a cube of 3D lattice with reflective boundary conditions. Figure~\ref{fig:Tks}(a) shows both $\langle T_k \rangle$ and $T_k$, that they almost overlap on each other. Figure~\ref{fig:Tks}(b) plots the relative difference, i.e., $|\langle T_k \rangle - T_k|/T_k$. It is clear that the relative difference is within 2\%, indicating that the approximation of $T_k$ by $\langle T_k \rangle$ is valid.

\section{\label{sec:fullcover} Derivation of the rescaled full cover-time distribution}

For non-compact random search process, the cover-time distribution can be estimated by Eq.~(\ref{eq:cover}), i.e.,
\begin{equation}
\begin{aligned}
P(\tau) = \frac{1}{N}\sum_{k,s}^{k\neq s} F_k(\tau) \left[ \prod_{i}^{i\notin \{k, s\}} \sum_{t}^{\tau-1} F_i(t) \right],
\end{aligned}
\label{eq:A:cover}
\end{equation}
where $F_k(t)$ is the probability density of the FPT to site $k$. According to the analysis in Sec.~\ref{sec:theory}, $F_k(t)$ is an exponential function in the large FPT limit, replacing $T_k$ by $\langle T_k \rangle$, one has
$$
F_k(t) \cong \frac{1}{\langle T_k \rangle} \exp\left(-\frac{t}{\langle T_k \rangle}\right).
$$
Since $\tau \gg 1$, $\tau - 1 \cong \tau$, then the summation $\sum F_k(t)$ can be approximated as
\begin{equation}
\sum_{t=1}^{\tau} F_k(t) \cong 1 - \exp\left(-\frac{\tau}{\langle T_k \rangle}\right).
\label{eq:A:mid0}
\end{equation}
Substituting the above equation into Eq.~(\ref{eq:A:cover}), and considering $N \gg 1$, the term in the square bracket can be approximated as
\begin{equation}
\begin{aligned}
\prod_{i} \sum_{t}^{\tau} F_i(t)
\cong& \prod_{i} \left(1 - \exp\left(-\frac{\tau}{\langle T_i \rangle} \right) \right) \\
\approx & 1-\sum_i \exp\left(-\frac{\tau}{\langle T_i \rangle}\right)+ \\
& \frac{1}{2}\sum_{i\neq j}\exp\left(-\frac{\tau}{\langle T_i \rangle}-\frac{\tau}{\langle T_j \rangle}\right) - \cdots .
\end{aligned}
\label{eq:A:mid1}
\end{equation}
Note that the summation in the last term excludes the case $i=j$, as $i$ appears only once in the product before the expansion. This also occurs in higher order terms, where the terms with identical indices are excluded. However, the number of those terms is about $1/N$ of all the terms, and since $\tau$ is typically much larger than $\langle T_i \rangle$ that the most dominant term is the lowest order term, the above expression can be approximated by
\begin{equation}
\prod_{i} \sum_{t}^{\tau} F_i(t)
\cong \exp\left( - \sum_{i=1}^{N} \exp\left(- \frac{\tau}{\langle T_i \rangle}\right) \right).
\label{eq:A:mid2}
\end{equation}

The above equation can be used to define a rescaling transformation to obtain the rescaled cover time $\chi$ from the original cover time $\tau$
\begin{equation}
\exp(-\chi) = \sum_i \exp\left(- \frac{\tau}{\langle T_i \rangle} \right),
\label{eq:A:mid3}
\end{equation}
or
\begin{equation}
\chi= -\ln \sum_i e^{- \frac{\tau}{\langle T_i \rangle} }.
\label{eq:A:xstar}
\end{equation}
Thus for a given set of $\{\langle T_i \rangle\}$, the rescaling relation between $\chi$ and $\tau$ is a monotonous deterministic function. Substituting Eqs. (\ref{eq:A:mid2}) and (\ref{eq:A:mid3}) back to Eq. (\ref{eq:A:cover}), and note that from Eq. (\ref{eq:A:mid3}), $${\rm d}\chi/{\rm d}\tau = \left[\sum_{k}^{N}\frac{1}{\langle T_k \rangle} \exp\left(-\frac{\tau}{\langle T_k \rangle}\right)\right]/\exp(-\chi),$$ one has
\begin{equation}
\begin{aligned}
P(\tau) & \cong \frac{1}{N}\sum_{k,s}^{k\neq s} F_k(\tau) \exp[-\exp(-\chi)] \\
& = \exp[-\exp(-\chi)] \sum_{k}^{N}\frac{1}{\langle T_k \rangle} \exp\left(-\frac{\tau}{\langle T_k \rangle}\right) \\
& = \exp[-\exp(-\chi)] \exp(-\chi) {\rm d}\chi/{\rm d}\tau.
\end{aligned}
\label{eq:A:mid4}
\end{equation}
Since $P(\tau){\rm d}\tau = P(\chi) {\rm d}\chi$, one has
\begin{equation}
P(\chi) \equiv \exp(-\chi-\exp(-\chi)),
\label{eq:A:mid5}
\end{equation}
which is the rescaled distribution function.

\section{ \label{secA:partial} Derivation of the rescaled partial cover-time distribution}

In Sec.~\ref{sec:theory}, the partial cover-time distribution is given by
\begin{equation}
P_m(\tau) = \sum_{\{I_m\}}
P(\tau | \notin \{I_m\}) \, Q(\tau , \{I_m\}),
\label{eq:A:partial0}
\end{equation}
where $\{I_m\}=\{i_1,i_2,\cdots,i_m\}$ is a particular set of $m$ sites, and $P(\tau| \notin \{I_m\})$ is the time distribution of covering all the other $N-m$ sites, no matter it has been covered or not for the sites in $\{I_m\}$. The function $Q(\tau , \{I_m\})$ is the probability that the walker does not visit any of the sites in $\{I_m\}$ during time $\tau$. After enumerating all possible configurations of $\{I_m\}$ by the summation $\sum_{\{I_m\}}$, the partial cover-time distribution $P_m(\tau)$ is then obtained.

$P(\tau | \notin \{I_m\})$ is the distribution for a random cover process. Using Eqs. (\ref{eq:A:mid3}) and (\ref{eq:A:mid5}), it can be written as
\begin{equation}
\begin{aligned}
&P(\tau | \notin \{I_m\})= P(\chi | \notin\{I_m\}) \frac{\mathrm{d}\chi}{\mathrm{d}\tau} \\=& \exp[-\exp(-\chi)]\exp(-\chi)\frac{\mathrm{d}\chi}{\mathrm{d}\tau}\bigg|_{i\notin \{I_m\}}.
\end{aligned}
\label{eq:P6}
\end{equation}
Note that
$$\exp(-\chi)|_{i\notin \{I_m\}} = \sum_i^{i\notin \{I_m\}} \exp\left(-\frac{\tau}{\langle T_i \rangle}\right).$$
Since $m \ll N$, the summation of $N-m$ on the right hand side will be close to the summation over all the $N$ terms, where the difference is in the order of $m/N$. Therefore, $\exp(-\chi)|_{i\notin \{I_m\}} \cong \exp(-\chi)$.

Thus
\begin{equation}
\begin{aligned}
P(\tau|\notin\{I_m\})=&P(\tau)=P(\chi)\frac{\mathrm{d}\chi}{\mathrm{d}\tau} \\
=& \exp\left(-\chi - e^{-\chi}\right)\frac{\mathrm{d}\chi}{\mathrm{d}\tau},
\end{aligned}
\label{eq:P7}
\end{equation}
which is almost independent to $\{I_m\}$.

If the unvisited sites are uncorrelated, the probability $Q$ can be explicitly obtained,
\begin{equation}
Q(\tau , \{I_m\}) = \prod_i^{\{I_m\}} \left(1-\sum_t^{\tau} F_i(t) \right),
\label{eq:A:mid4}
\end{equation}
where $F_i(t)$ is the probability density of the FPT to site $i$. According to Eq.~(\ref{eq:A:mid0}), $Q(\tau , \{I_m\})$ can be expressed as
\begin{equation}
Q(\tau , \{I_m\}) = \prod_{i}^{\{I_m\}} \exp\left(-\frac{\tau}{\langle T_i \rangle}\right),
\label{eq:A:Q5}
\end{equation}

Equation~(\ref{eq:A:partial0}) for $P_m(\tau)$ can then be represented as
$$
P_m(\tau) \sim \exp\left(-\chi - e^{-\chi}\right) \left[\sum_{\{I_m\}}\prod_i^{\{I_m\}} \exp\left(-\frac{\tau}{\langle T_i \rangle}\right)\right] \frac{\mathrm{d}\chi}{\mathrm{d}\tau}.
$$
The summation in the square bracket is over all possible combinations of $m$ different sites out of the total $N$ sites, in total of
$$\bigg(\begin{array}{c}
 N \\
 m
\end{array}
\bigg)=N\times \cdots \times (N-m+1)/m!$$
terms. Note that the number of terms stated above excludes those with identical indices, which are in general only a small fraction, i.e., in an order of $m/N$. For $m \ll N$, their contributions can be neglected, and the summation in the square bracket can be approximated as
\begin{equation}
\frac{1}{m!}\left[\sum_i\exp\left(-\frac{\tau}{\langle T_i \rangle}\right)\right]^m=\frac{1}{m!}\exp(-m \chi).
\label{eq:A:A1}
\end{equation}
Then, the uncorrelated partial cover-time distribution is
\begin{equation}
P_m(\chi) = \frac{1}{m!} \exp\left(-(m+1)\chi-\exp(-\chi)\right),
\label{eq:A:partial}
\end{equation}
and the normalization condition is satisfied naturally.

When the unvisited sites have non-negligible but still small correlation, the approximation Eq.~(\ref{eq:A:A1}) may be inaccurate. As a consequence, the value of $m$ in Eq.~(\ref{eq:A:partial}) has to be corrected to a smaller value $m^*$, while the same form Eq.~(\ref{eq:A:partial}) of the partial cover time may still be valid.

\bibliography{covertime}

\begin{thebibliography}{77}
\bibitem [{Ref()}]{Ref3D}%
  \BibitemOpen
  \bibinfo {note} {Starting from a given inner lattice site $s$, the walker
  moves to any of the six neighboring sites with identical probability $1/6$.
  If $s$ is on the face and next step the walker moves out of the domain, it
  will then be reflected back by the face, say, to site $i$. Or equivalently,
  $\omega_{i \leftarrow s}=1/3$, and the transfer probabilities to the other
  four neighbors on the face are all $1/6$. If $s$ is on the edge, it has four
  neighbors, the two also on the edge each has a transfer probability $1/6$,
  while the other two neighbors each has a probability $1/3$. If $s$ is at the
  corner, then it only has three neighbors, each has a transfer probability of
  $1/3$.}\BibitemShut {Stop}%
\end{thebibliography}

\end{document}